\documentclass[fleqn,12pt,twoside]{article}
\usepackage{espcrc1}
\usepackage{epsfig}

\usepackage{graphicx}

\newcommand{\AmS}{{\protect\the\textfont2
  A\kern-.1667em\lower.5ex\hbox{M}\kern-.125emS}}
\def\bbox#1{\mbox{\boldmath $#1$}}

\title{A meson-exchange $\pi N$ model up to   energies
$\sqrt{s} \le 2.0$  GeV}

\author{Shin Nan Yang\address{Department of Physics and Center for Theoretical Sciences \\
        National Taiwan University, Taipei 10617, Taiwan}%
        \thanks{This work is supported in part by the National Science Council of ROC under grant No.
        NSC94-2112-M002-025.},
        Guan Yeu Chen\addressmark,
        and
        S.S. Kamalov\address{Bogoliubov Laboratory for Theoretical Physics, JINR,\\
        Dubna, 141980 Moscow Region, Russia}}

\begin{document}

\maketitle

\begin{abstract}
A meson-exchange $\pi N$ model, previously constructed using
three-dimensional reduction scheme of the Bethe-Salpeter equation
for a model Lagrangian involving $\pi$, $\eta$, $N$, $\Delta$,
$\rho$, and $\sigma$ fields, is extended to energies up to 2 GeV
by including the $\eta N$ channel and all the four stars $\pi N$
resonances up to the $F-$waves. The effects of other $2\pi$
channels are taken into account phenomenologically. The extended
model gives an excellent fit to both $\pi N$ phase shifts and
inelasticity parameters in all channels up to the $F-$waves.
However, a few of the extracted resonance parameters differ
considerably from the PDG values.
\end{abstract}

\section{INTRODUCTION}

Pion-nucleon scattering is one of the main sources of information
for the baryon spectrum. In addition, it also plays a fundamental
role in the description of nuclear dynamics for which the $\pi N$
off-shell amplitude serves as the basic input to most of the
existing nuclear calculations at intermediate energies. Knowledge
about the off-shell $\pi N$ amplitude is also essential in
interpreting the experiments performed at the intermediate-energy
electron accelerators in order to unravel the internal structure of
these hadrons \cite{yang85,lee91,KY99}. It is hence important to
have a sound theoretical description of the $\pi N$ interaction.

It is commonly accepted that Quantum Chromodynamics (QCD) is the
fundamental theory of the strong interaction. However, due to the
confinement problem, it is still practically impossible  to derive
the $\pi N$ interaction directly from QCD. On the other hand, models
based on meson-exchange (MEX) pictures  have been very successful in
describing the $NN$ scattering. Over the past decade, MEX approach
has also been applied by several groups
\cite{pj,gs,juelich,lahiff99,pt00,hung01} to construct models for
$\pi N$ scattering.

In  previous works we have constructed several MEX $\pi N$ models
within the three-dimensional reduction scheme of the
Bethe-Salpeter equation \cite{lee91,hung01} and investigated their
sensitivity with respect to various three-dimensional reduction
schemes. The model Lagrangian included only $\pi, N, \Delta,
\rho,$ and $\sigma$ fields. It was found that all the resulting
meson-exchange models can yield similarly good descriptions of
$\pi N$ scattering data up to $400$ MeV. The model obtained with
the Cooper-Jennings reduction scheme \cite{CJ} was recently
extended up to a c.m. energy of 2 GeV in the $S_{11}$ channel by
including the $\eta N$ channel and a set of higher $S_{11}$
resonances \cite{chen03}. The effects of the other $\pi\pi N$
channels like the $\sigma N$, $\rho N$, and $\pi\Delta$, instead
of including them  directly in the coupled-channels calculation,
were taken into account by introducing a phenomenological term in
the resonance propagators. An excellent fit to the t-matrix in
both $\pi N$ and $\eta N$ channels was obtained. Here we further
extend the model to other higher partial waves up to the
$F-$waves.

\section{MESON-EXCHANGE MODEL FOR $\pi N$ SCATTERING}

The MEX  $\pi N$ model we previously constructed was obtained by
using a three-dimensional reduction scheme of the Bethe-Salpeter
equation for a model Lagrangian involving $\pi, N, \Delta, \rho,$
and $\sigma$ fields. Details  can be found in Ref. \cite{hung01}.

As the energy increases, two-pion channels like $\sigma N, \eta N,
\pi\Delta, \rho N$ as well as a non-resonant continuum of $\pi \pi
N$ states become increasingly important, and at the same time more
and more nucleon resonances appear as intermediate states. In
Ref.~\cite{chen03} the $\pi N$ model constructed in \cite{hung01}
was  extended for the $S_{11}$ partial wave by explicitly coupling
the $\pi$, $\eta$ and $\pi \pi$ channels and including the
couplings with higher baryon resonances. For example, in the case
when there is only one resonance $R$ contributing, the Hilbert
space was enlarged to include a  bare $S_{11}$ resonance $R$ which
acquires a width by its coupling with the $\pi N$ and $\eta N$
channels through the Lagrangian
\begin{eqnarray}
{\cal{L_I}}= ig^{(0)}_{\pi NR}\bar R\bbox{\tau}N\cdot\bbox{\pi} +
ig^{(0)}_{\eta NR}\bar R N\eta + h.c., \label{lagr}
\end{eqnarray}
where $N, R, \bbox{\pi},$ and $\eta$ denote the field operators
for the nucleon, bare $R$, pion and eta meson, respectively. Then
the full $t$-matrix can be written as a system of coupled
equations,
\begin{eqnarray}
t_{ij}(E)= v_{ij}(E)+\sum_k  v_{ik}(E)\,g_k(E)\, t_{kj}(E)\,,
\label{t_ij}
\end{eqnarray}
where $i$ and $j$ denote the $\pi$, or $\eta$ channel and $E=W$ is
the total center mass energy.

In general, the potential $ v_{ij}$ is a sum of non-resonant
$(v^B_{ij})$ and bare resonance $(v^R_{ij})$ terms,
\begin{eqnarray}
v_{ij}(E)=  v^B_{ij}(E)+ v^R_{ij}(E)\,. \label{v_ij}
\end{eqnarray}
The non-resonant term $v^B_{\pi\pi}$ for the $\pi N$ elastic
channel contains contributions from the $s$- and $u$-channel,
pseudovector Born terms and $t$-channel contributions with
$\omega$, $\rho$, and $\sigma$ exchange. The parameters in
$v^B_{\pi\pi}$ are fixed from the analysis of the pion scattering
phase shifts for the $S-$ and $P-$waves at low energies ($W<1300$
MeV) \cite{hung01}. In channels involving $\eta$, $v^B_{i\eta}$ is
taken to be zero since the $\eta NN$ coupling is very small.

The bare resonance contribution \begin{eqnarray} v^R_{ij}(E)=\frac
{h_{i R}^{(0)\dagger} h_{j R}^{(0)}}{E-M_R^{(0)}}, \label{vRbare}
\end{eqnarray}
where $h_{i R}^{(0)}$ and $M_R^{(0)}$ denote the bare vertex
operator for $\pi/\eta + N \rightarrow R$ and bare mass of $R$,
respectively, arises from the excitation and de-excitation of the
resonance $R$. The matrix elements of the potential $v^R_{ij}(E)$
can be symbolically expressed in the form
\begin{eqnarray}
v^R_{ij}(q,q';E)=\frac{f_i(\tilde
{\Lambda}_i,q;E)\,g_i^{(0)}\,g_j^{(0)}\,
f_j(\tilde{\Lambda}_j,q';E)}{E-M_R^{(0)}+
\frac{i}{2}\Gamma^R_{2\pi}(E)} \,, \label{v_R1}
\end{eqnarray}
where $q$ and $q'$ are the pion (or eta) momenta in the initial
and final states, and $g_{i/j}^{(0)}$ is the resonance vertex
couplings. As in \cite{hung01}, we associate with each external
line of the particle $\alpha$ in a Feynman diagram a covariant
form factor $F_\alpha =
[n_\alpha\Lambda^4_\alpha/(n_\alpha\Lambda^4_\alpha+(p^2_\alpha
-m^2_\alpha)^2)]^{n_\alpha}$, where $p_\alpha$, $m_\alpha$, and
$\Lambda_\alpha$ are the four-momentum, mass, and cut-off
parameter of particle $\alpha$, respectively, and $n_c=10$. As a
result, $f_i$ depends on the  product of three cut-off parameters.

In Eq. (\ref{v_R1}) we have included a phenomenological term
$\Gamma^R_{2\pi}(E)$ in the resonance propagator to account for
the $\pi\pi N$ decay channel. Therefore, our "bare" resonance
propagator already contains some renormalization or "dressing"
effects due to the coupling with the $\pi\pi N$ channel. With this
prescription we assume that any further non-resonant coupling
mechanisms with the $\pi\pi N$ channel are small. The form  of
$\Gamma^R_{2\pi}(E)$ can be found in \cite{chen03} and is
characterized by two parameters, a cut-off $\Lambda_R$ and the
$2\pi$ decay width at the resonance $\Gamma^{(0)R}_{2\pi}$.
Consequently, one isolated resonance will contain five free
parameters, $M^{(0)}_R, \Gamma^{(0)R}_{2\pi}, \Lambda_R,
\,g^{(0)}_i$ and $g^{(0)}_j$. The generalization of the coupled
channels model to the case of $N$ resonances with the same quantum
numbers is then given by
\begin{eqnarray}
v^R_{ij}(q,q';E)=\sum_{n=1}^{N} v^{R_n}_{ij}(q,q';E), \label{v_RN}
\end{eqnarray}
with free parameters for the bare masses, $2\pi$ widths, coupling
constants, and cut-off parameters for each resonance.

After solving the coupled channel equations, the next task is the
extraction of the physical (or "dressed") masses, partial widths,
and branching ratios of the resonances.  It is well-known this
procedure is definitely model dependent, because the background
and the resonance contributions can not be separated in a unique
way. In this work, we employ the procedure used in
Ref.~\cite{KY99} where, in the case of  pion-nucleon elastic
scattering with only one resonance contributing, the full t-matrix
is written as follows,
\begin{eqnarray}
t_{\pi N}(E)=t_{\pi N}^B(E) + t_{\pi
N}^{R}(E),\label{eq:tgammapi33}
\end{eqnarray}
where
\begin{eqnarray}
t_{\pi N}^B(E)=v_{\pi N}^B+v_{\pi N}^B\,g_0(E)\,t_{\pi N}(E)\,,
\hspace{1.5cm} t_{\pi N}^{R}(E)=v_{\pi N}^R+v_{\pi B}^R\,g_0(E)
\,t_{\pi N}(E). \label{DMT}
\end{eqnarray}
The "background" $t_{\pi N}^B$  includes contributions not only from
the background rescattering but also from intermediate resonance
excitation. This is compensated by the fact that the resonance
contribution $t_{\pi N}^{R}$ now contains only the terms that start
with the bare resonance vertex. In terms of self-energy and vertex
functions, one obtains the result \cite{Hsiao98}
\begin{eqnarray}
t_{\pi N}^{R}(E)=\frac{\bar h_{\pi R}(E) h_{\pi
R}^{(0)}(E)}{E-M_{R}^{(0)}(E) -\Sigma_{R}(E)} \, , \label{tR_DMT2}
\end{eqnarray}
where
\begin{eqnarray}
\bar h_{\pi R}(E)=(1+\,g_0(E) \,t_{\pi N}^{B}(E))h^{(0)\dagger}_{\pi
R}, \hspace{1.2cm} \Sigma^R_{1\pi}(E)= h_{\pi R}^{(0)}\,g_0\, \bar
h_{\pi R}(E). \label{Sigma_Delta}
\end{eqnarray}
$\bar h_{\pi R}(E)$ describes the dressed vertex of $R\rightarrow
\pi N$ \cite{KY99}. $\Sigma_{1\pi}^{R}$ is the self-energy of the
dressed $R$ arising from one-pion intermediate states and
$\Sigma_{R}(E)=\Sigma^R_{1\pi} + \Sigma^R_{2\pi}$ with
$\Sigma^R_{2\pi}(E)=2 i\Gamma^R_{2\pi}$.

The information about the physical mass and the total width of the
resonance $R$ are contained in the dressed resonance propagator
given in Eq.~(\ref{tR_DMT2}). The complex self-energy $\Sigma_R(E)$
leads to a shift from the real "bare" mass to a complex and
energy-dependent value. We characterize the resonance by
energy-independent parameter that is obtained by solving the
equation
\begin{eqnarray}
E-M_{R}^{(0)}-Re \, \Sigma_{R}(E)=0  \, . \label{Delta_mass1}
\end{eqnarray}
The solution of this equation, $E=M_R$, corresponds to the energy at
which the dressed propagator in Eq. (\ref{tR_DMT2}) is purely
imaginary. The "physical" or "dressed" mass and the width of the
resonance is then defined by,
\begin{eqnarray}\label{Mdressed}
{M}_{R}=M_{R}^{(0)}+Re\,\Sigma_{R}({M}_{R}), \hspace{2.0cm}
\Gamma_{R} =-2\,Im\,\Sigma_{R}(M_R). \label{Mass-width}
\end{eqnarray}

When there are N resonances contributing in the same channel, Eq.
(\ref{DMT}) can be generalized to take the form of
\begin{eqnarray}
t_{\pi N}(E)= t^{B}_{\pi N}(E)+ \sum_{i=1}^{N}  t^{R_i}_{\pi
N}(E)\,. \label{T_piN}
\end{eqnarray}
The contribution from each resonance $R_i$ can be expressed in terms
of the bare $h^{(0)}_{\pi  R_i}$ and dressed $\bar h_{\pi  R_i}$
vertex operators as well as the resonance self energy derived from
one-pion $\Sigma_{1\pi}^{R_i}(E)$ and two-pion
$\Sigma_{2\pi}^{R_i}(E)$ channels, that is
\begin{eqnarray}
t_{\pi N}^{R_i}(E)=\frac{\bar h_{\pi R_i}(E) h_{\pi R_i
}^{(0)}(E)}{E-M_{R_i}^{(0)}-\Sigma_{1\pi}^{R_i}(E)-\Sigma_{2\pi}^{R_i}(E)}\,,
\label{tRidressed}
\end{eqnarray}
where $M_{R_i}^{(0)}$ is the bare mass of $R_i$. The vertices for
resonance excitation are obtained, in analogous to Eqs.
(\ref{DMT}-\ref{Sigma_Delta}), from the following two equations:
\begin{eqnarray}
&&\bar h_{\pi R_i}(E)=(1+\,g_0(E)
\,t_{\pi N}^{B_i}(E))h_{\pi R_i}^{(0)\dagger}  \label{hRi}\\
&&t_{\pi N}^{B_i}(E)=v^B_{i}(E)+v^B_i(E)\,g_0(E)\,t_{\pi
N}^{B_i}(E)\,, \label{DMT_bcr}
\end{eqnarray}
where $v^B_i(E)=v_{\pi N}^B+\sum_{j\neq i}^{N}\,v_{\pi N}^{R_j}(E)$.
The one-pion self-energies corresponding to Eq.~(\ref{Sigma_Delta}),
is $\Sigma_{1\pi}^{R_i}=h_{\pi R_i}^{(0)}\, g_0 \,\bar h_{\pi
R_i}^{\dagger}$. We wish to emphasize that  in the formulation we
present above, namely in Eqs. (\ref{T_piN}-\ref{DMT_bcr}), the N
resonances are treated in a completely symmetrical way and the
self-energy and the dressing of any resonance receive contributions
from all other resonances.
\section{RESULTS AND DISCUSSIONS}

In Fig. 1, we compare our results for the real and imaginary parts
of the t-matrix in some selected channels in $S-, P-, D-$ and
$F-$waves up to 2 GeV c.m. energy with the experimental data as
obtained in the SAID partial wave analysis \cite{SAID04}. One sees
that we are able to describe the data very well.
\begin{figure}[!htb]
\centering \epsfig{file=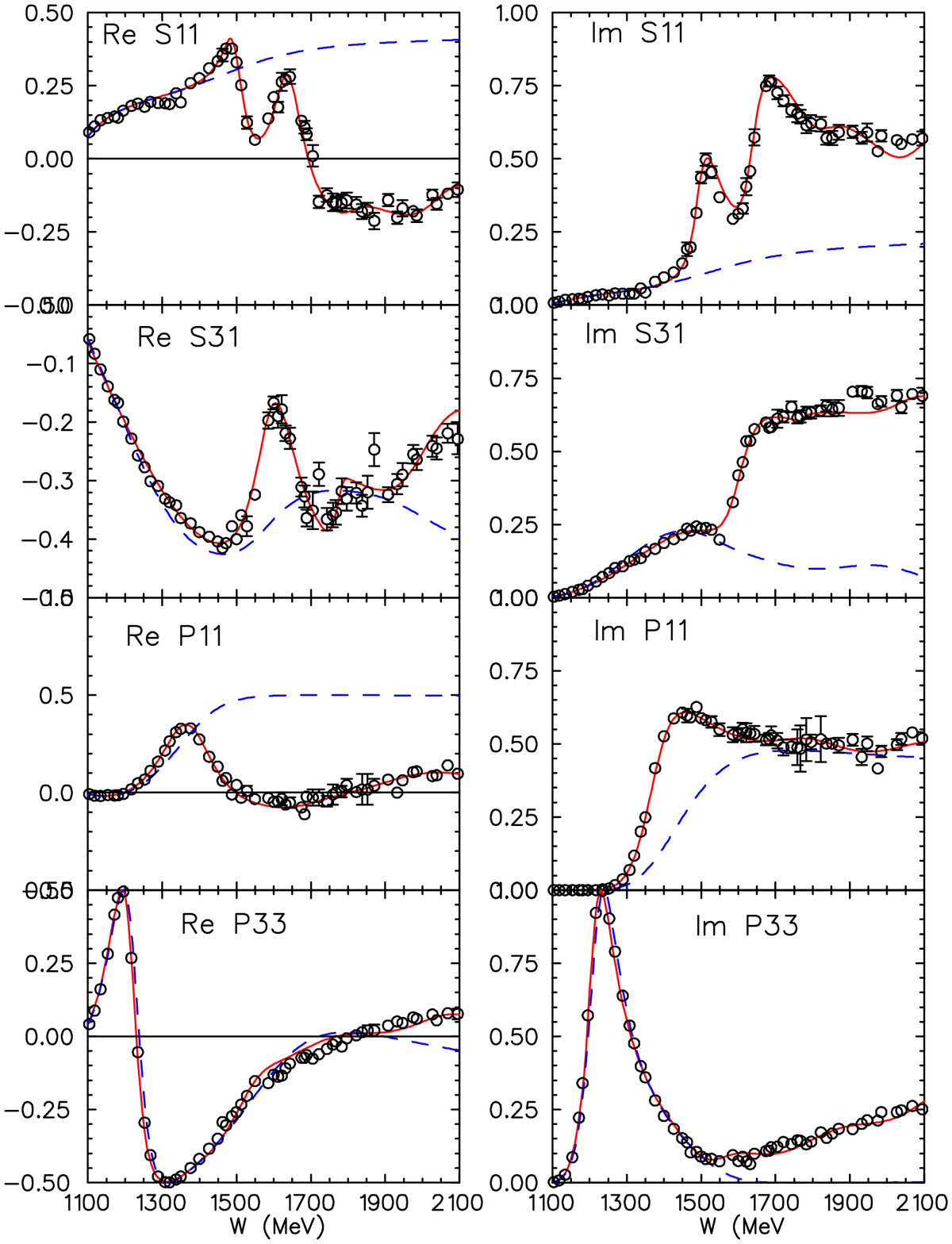, width=6cm}
\epsfig{file=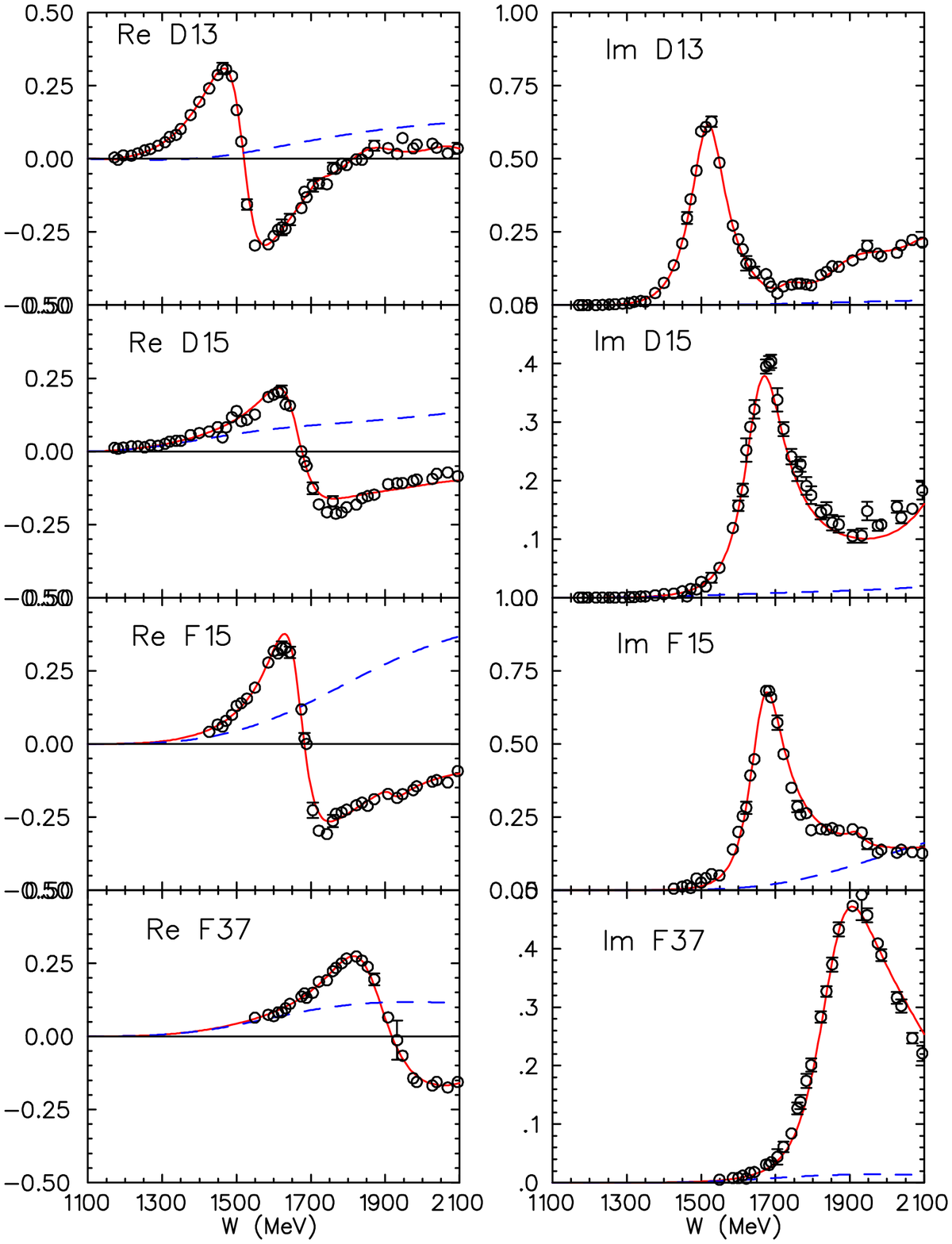, width=6cm} \caption{\small The best fit
of the real and imaginary parts of the $\pi N$ scattering t-matrix
using dynamical MEX model (solid curves). The dashed curves give
the background contribution. Experimental data are the results of
the partial wave analysis from Ref.~\cite{SAID04}. }
\end{figure}

 The bare and physical resonance masses, and widths
extracted according to Eq. (\ref{Mass-width}) are presented in
Table 1. Even though our model describes the data for t-matrix
well as seen in Fig. 1, the resonance properties we extract as
given in Table 1 do show several differences when compared with
PDG values \cite{PDG06}. The most notable ones are that (1). we
require two resonances not listed in PDG: $S_{11}(1878)$ and
$D_{13}(1946)$; (2). the masses and widths we obtain for the 2nd
and 3rd resonances in $S_{31}$ and $P_{11}$  deviates
substantially from the PDG values. The PDG values for
$(M_R,\Gamma_R)$ for these resonances are $S_{31}: (1900\pm
50,190\pm 50), (2150\pm 100, 200\pm 100)$ and $P_{11}: (1710\pm
30, 180\pm 100), (2125\pm 75, 260\pm 100)$; (3). the width we
obtain for  $F_{15}(2000)$ is only 58 MeV which is much smaller
that the PDG value of $490\pm 310$ MeV.

\begin{table}[!htb]
\centering \setlength{\textwidth}{50mm} \caption{Bare $M_R^{(0)}$
and physical $M_R$ resonance masses and total width in MeV.}
\begin{tabular}{|l|ccc|ccc|ccc|}
\hline
       & & 1st res. & &  & 2nd res. &  &  & 3rd res.  & \\
\hline
 $N^*$ & $M_R^{(0)}$ & $M_R$ & $\Gamma$  &  $M_R^{(0)}$ & $M_R$ & $\Gamma$
 & $M_R^{(0)}$ & $M_R$ & $\Gamma$
 \\
\hline
$S_{11}$  & 1559 & 1520 & 130  & 1727 & 1678 & 200  & 1803 & 1878 & 508  \\
$S_{31}$  & 1654 & 1616 & 160  & 1796 & 1770 & 430  & 2118 & 1942 & 416  \\
$P_{11}$  & 1612 & 1418 & 436  & 1798 & 1803 & 508  & 2196 & 2247 & 1020 \\
$P_{33}$  & 1425 & 1233 & 132  & 1575 & 1562 & 216  & 1856 & 1827 & 834  \\
$D_{13}$  & 1590 & 1520 & 94   & 1753 & 1747 & 156  & 1972 & 1946 & 494  \\
$D_{33}$  & 1690 & 1650 & 260  & 2100 & 2092 & 310  & --- & --- & ---  \\
$D_{15}$  & 1710 & 1670 & 154  & 2300 & 2286 & 532  & ---  & ---  & ---  \\
$F_{15}$  & 1748 & 1687 & 156  & 1928 & 1926 & 58   & ---  & ---  & ---  \\
$F_{37}$  & 1974 & 1916 & 338  & ---  & --- & ---  & ---  & ---  & ---  \\
\hline
\end{tabular}
\end{table}

\section{SUMMARY}

We have extended  our previously constructed meson-exchange $\pi
N$ model to energies up to 2 GeV by including the $\eta N$ channel
and all the four stars $\pi N$ resonances up to the $F-$waves. The
effects of other $2\pi$ channels are taken into account
phenomenologically. We have treated, in any given channel, all the
contributing resonances in a completely symmetrical manner such
that every resonance is dressed by the presence of all other
resonances. The extended model gives an excellent fit to both $\pi
N$ phase shifts and inelasticity parameters in all channels up to
the $F-$waves. However, a few of the resonance parameters differ
substantially from the PDG values. This $\pi N$ model will be
applied to evaluate the contribution of the pion cloud to the
photopion reactions up to 2 GeV c.m. energy as was done in Ref.
\cite{KY99} so that the photoexcitation strengths of all
resonances below 2 GeV can be reliably extracted.

\end{document}